# An Automated Vehicle (AV) like Me?
# The Impact of Personality Similarities and Differences between Humans and AVs

**Qiaoning Zhang**[1]
MAVRIC
University of Michigan
qiaoning@umich.edu

**Connor Esterwood**[2]
MAVRIC
University of Michigan
cte@umich.edu

**X. Jessie Yang**
MAVRIC
University of Michigan
xijyang@umich.edu

**Lionel P. Robert Jr.**
MAVRIC
University of Michigan
lprobert@umich.edu

## Abstract

To better understand the impacts of similarities and dissimilarities in human and AV personalities we conducted an experimental study with 443 individuals. Generally, similarities in human and AV personalities led to a higher perception of AV safety only when both were high in specific personality traits. Dissimilarities in human and AV personalities also yielded a higher perception of AV safety, but only when the AV was higher than the human in a particular personality trait.

## Introduction

Autonomous vehicles (AVs) are an artificial intelligence (AI)-enabled service robots. AVs are expected to provide more fuel-efficient and safer driving (Chen, Wang, and Meng 2019; Katrakazas et al. 2015; Young and Stanton 2004; Eby et al. 2016; Robert 2019). Yet, there are doubts about whether individuals will adopt AVs (Du et al. 2019).

One solution to promoting the acceptance of AVs is to design them to have a similar personality as their human riders. Research on human-to-human interactions has found that humans often prefer interacting with other humans with a similar personality (Byrne and Griffitt 1969). However, the literature on human and robot personalities has found mixed results. Some studies have found that similarities in human and robot personalities led to positive human-robot interactions(Aly and Tapus 2016; Tapus and Mataric 2008). Others have found that dissimilarities in personalities led to positive interactions (Lee et al. 2006).

To better understand the impacts of similarities and dissimilarities in human and AV personalities we conducted a study employing a nationwide survey of 443 individuals. This study examined the impacts of similarities and dissimilarities in human and AV personalities as they relate to the Big Five personality traits. Generally, similarities in human and robot personalities increased perceptions when both the human and AV were high in particular personality traits. Dissimilarities in human and robot personalities also yielded increases in perceptions of AV safety, but only when the AV was higher than the human in a personality trait. The positive impacts of both similarities and dissimilarities were limited to agreeableness, conscientiousness and emotional stability. No such effects were found for extroversion or openness to experience. Finally, there was a moderation effect involving the experimental condition on the relationship between conscientiousness and AV safety.

## Background
### Personality and the Big Five

Personality can be used to predict human attitudes, emotions and behaviors (Robert 2018). Personality is used as a label to describe traits that represent an individual's predisposition toward a behavior or object. Personality is now a core construct in understanding human-robot interactions (for a review see Robert 2018).

The Big Five is the most popular set of personality traits in social science in general and in the study of human-robot interaction specifically (Robert 2018). The Big Five include: (1) extroversion, defined as being sociable, gregarious, and ambitious; (2) agreeableness, defined as being kind, considerate, likeable, and cooperative (Graziano and Eisenberg 1997); (3) conscientiousness, which reflects self-control and a need for achievement and order; (4) emotional stability, characterized by being well-adjusted, emotionally stable and secure; and (5) openness to experience, which is represented by flexibility of thought and tolerance of new ideas (Costa Jr, McCrae, and Dye 1991; Devaraj, Easley, and Crant 2008; Graziano and Eisenberg 1997).

### Personality and Human-Robot interaction

The literature on human-robot personality can be grouped into three sets. First, several authors have found that similarity in human and robot personality can lead to positive interactions. These studies are based on the underlying logic that birds of a feather flock together(Byrne and Griffitt 1969). For example, several studies have found that humans prefer interacting with robots that have their own personality over robots who have a different personality (Aly and Tapus 2016; Tapus and Mataric 2008).

Second, prior research has also found that dissimilarity in human and robot personality can lead to positive interactions. These studies were based on the underlying logic that opposites attract. This assertion has also been supported by several studies which found that humans preferred interacting with robots that had a different personality from theirs (Celiktutan and Gunes 2015; Lee et al. 2006).

Finally, another view is that the impacts of human-robot similarity or dissimilarity depends heavily on a given context (Joosse et al. 2013). For example, Joosse et al. (2013) found that the relationship between similar personalities and human preference for a robot were moderated by task type.

Taken as a whole, the literature on the impacts of personality similarity/dissimilarity between humans and robots has not found consistent results. In addition, little effort has been made to examine their impacts as they relate to human and AV interactions.

## Methodology

This study employed an experimental design and was approved by the institutional review board.

### Respondents and Survey Instruments

A total of 443 U.S.-licensed drivers (mean age = 47.2 years, standard deviation [SD] = 15.8 years) participated. The sample was selected to represent the typical U.S. driving population based on statistics provided by the U.S. Department of Transportation and AAA Foundation(Triplett et al. 2016). This was done to ensure our sample represented the range in age, gender, ethnicity, and geographic regions of the United States. Qualtrics was hired to recruit the participants, who were paid.

### Procedure

Step 1, participants were required to fill out a consent form. Step 2, participants completed a survey asking for their demographic information to determine whether they qualified to participate in the study. Step 3, participants completed a survey that measured their personality. Step 4, each participant was required to watch four videos of an AV driving. Each video placed the participant in the front seat of an AV while it drove (see Figure 1). The four videos manipulated the AV driving behavior (i.e. normal or aggressive driving) and the weather conditions (i.e. sunny or snowy). The experimental study employed a 2 x 2 within-subjects design where each participant was randomly assigned to a particular video order to counterbalance any potential learning effects. Step 5, after each video, participants completed a survey to capture their perceptions of AV safety and the AV personality. Steps 4 and 5 were repeated for all four videos. Step 6, the participants were debriefed and paid.

### Measurements

#### Independent Variables

**Human-AV Interaction Conditions:** The "human-AV interaction variable" had four conditions based on the AV driving behaviors and the weather condition. Each factor had two levels representing four conditions, including: sunny and normal AV driving condition, sunny and aggressive AV driving condition, snowy and normal AV driving condition, and snowy and aggressive AV driving condition.

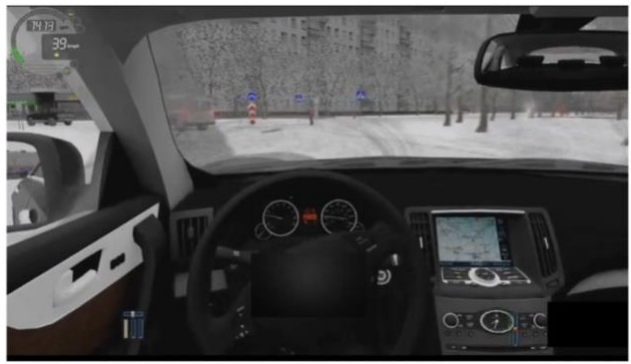

Figure 1: An AV drives in snowy weather.

**Personality Similarity and Dissimilarity:** The participant and AV personality questionnaires measured the Big Five personality traits: extroversion, agreeableness, conscientiousness, emotional stability, and openness to experience. Respondents were required to rate themselves and the AVs on a Ten-item Personality Inventory (TIPI) using a 7-point Likert scale (1: disagree strongly; 7: agree strongly) (Gosling, Rentfrow, and Swann Jr 2003).

The Big Five personality traits scores were divided into two groups consisting of high or low scores based on their means. Scores above the mean were classified as high and those below the mean were classified as low. This yielded a set of high and low personality categories for the participant as well as for the AV per treatment condition. The similarity and dissimilarity personality measures were as follows:

**Low Similarity**: Participant and the AV had similar personalities but both personality scores were low.
**High Similarity**: Participant and the AV had similar personalities and both personality scores were high.
**Low Dissimilarity**: Participant and the AV had dissimilar personalities in that the participant had a low personality score while the AV had a high personality score.
**High Dissimilarity**: Participant and the AV had dissimilar personalities in that the participant had a high personality score while the AV had a low personality score.

#### Dependent Variable

**Safety:** Safety was measured using a 10-item questionnaire(Hayes et al. 1998). All the items were rated on 5-point Likert scales (1: strongly disagree; 5: strongly agree).

## Results

The reliability of safety was 0.97, well above 0.70. A mixed liner model was used to analyze the data, with each personality having four categorical levels: high/low similarities and high/low dissimilarities between the participant and the AV. Safety was the dependent variable. Table 1 shows the results summary.

### Extroversion

Extroversion was not significant (F=0.632, p=0.595). However, the low dissimilarity had the lowest mean (see Table 1).

**Agreeableness**
The main effect of agreeableness similarities/dissimilarities on safety was significant (F=6.264, p<0.001). Post-hoc comparisons indicated that the low dissimilarity produced lower safety compared to low similarity and high dissimilarity (low dissimilarity vs. low similarity, p=0.002; low dissimilarity vs. high dissimilarity, p=0.002). Also, high similarity in agreeableness led to higher safety perception than high dissimilarity (p=0.002).

**Conscientiousness**
Safety perception was significantly influenced by conscientiousness similarities/dissimilarities (F=10.040, p<0.001). Post-hoc comparisons revealed that high dissimilarity had the lowest safety rating (high dissimilarity vs. low similarity, p=0.010; high dissimilarity vs. high similarity, p<0.001; high dissimilarity vs. low dissimilarity, p<0.001). Low dissimilarity led to the highest safety perception (low dissimilarity vs. low similarity, p=0.001; low dissimilarity vs. high similarity, p=0.031; low dissimilarity vs. high dissimilarity, p<0.001).

**Emotional Stability**
There is a significant effect of emotional stability on perceptions of AV safety (F=4.921, p=0.002). Post-hoc comparisons indicated that participants gave a higher safety rating when there was a high similarity or low dissimilarity than low similarity or high dissimilarity (high similarity vs. low similarity: p=0.045; high similarity vs. high dissimilarity: p=0.002; low dissimilarity vs. low similarity: p=0.004; low dissimilarity vs. high dissimilarity: p=0.044).

**Openness to Experience**
Openness to experience was not significant (F=0.897, p=0.442).

**Moderation Effects**
There was a significant moderation effect of human-AV interaction condition on the relationship between similar/dissimilar personality in conscientiousness (F=3.70, p<.001). Figure 2 displays the two-way interaction of the relationship. AV safety was generally higher for participants high in conscientiousness when the AV drove non-aggressively in sunny weather.

**Summary of the Results**
This paper has three main findings. First, there were no impacts associated with human-AV personality similarities/dissimilarities for extroversion and openness to experience. Second, for agreeableness, conscientiousness, and emotional stability, high similarity and/or low dissimilarity produced the highest perception of AV safety. Finally, there was a moderation effect associated with conscientiousness on safety. Generally, sunny weather and non-aggressive driving produced the highest perceptions of AV safety for people who are high conscientiousness.

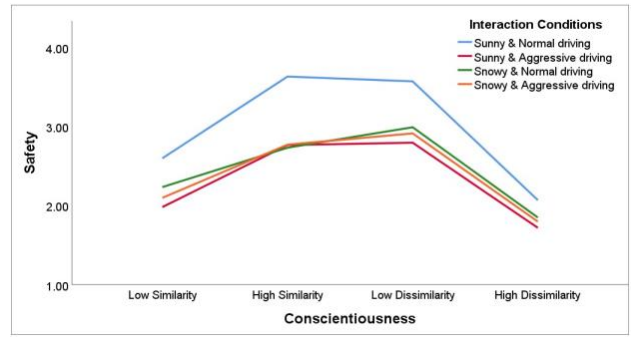

Figure 2: Interaction Effects

## Discussion

This study helps to explain the mixed results in the prior literature. First, this study explains when personality similarity could be beneficial. Our findings indicate that similarity led to higher perceptions of safety when both the AV and human had high scores on personality traits such as agreeableness, conscientiousness, and emotional stability. Our paper extends prior literature by highlighting the importance of high versus low personality scores for the impacts of similarity.

Second, this paper highlights when personality dissimilarity can also be good. In our study, dissimilarity was only good when perceptions of an AV's personality were higher than the human's personality with regard to agreeableness, conscientiousness, or emotional stability. Alternatively, when an AV's personality was perceived as lower than the human's personality in these traits, dissimilarity was likely to lead to lower perceived AV safety. Our paper extends prior literature by showing when dissimilarity is likely to lead to positive outcomes.

Finally, this paper provides some evidence of the importance of context. Our results support this conclusion by demonstrating the moderation effect of the experimental condition on the relationship between personality similarities/dissimilarities and AV safety.

Table 1: Safety rating summary

| Personality dimension | Independent variable | Safety means | Differences |
|---|---|---|---|
| Extraversion (F=0.632, p=0.595) | High dissimilarity | 2.63 | None |
| | High similarity | 2.62 | |
| | Low dissimilarity | 2.54 | |
| | Low similarity | 2.52 | |
| Agreeableness (F=6.264, p<0.001) | Low dissimilarity | 2.76 | Low similarity vs. Low dissimilarity, p=0.002 |
| | High similarity | 2.65 | High similarity vs. High dissimilarity, p=0.002 |
| | Low similarity | 2.50 | Low dissimilarity vs. High dissimilarity, p=0.002 |
| | High dissimilarity | 2.40 | |
| Conscientiousness (F=10.040, p<0.001) | Low dissimilarity | 2.81 | Low similarity vs. Low dissimilarity, p<0.001 |
| | High similarity | 2.64 | Low similarity vs. High dissimilarity, p=0.013 |
| | Low similarity | 2.55 | High similarity vs. High dissimilarity, p<0.001 |
| | High dissimilarity | 2.32 | Low dissimilarity vs. High dissimilarity, p<0.001 |
| Emotional Stability (F=4.921, p=0.002) | High similarity | 2.70 | Low similarity vs. High similarity, p=0.045 |
| | Low dissimilarity | 2.69 | Low similarity vs. low dissimilarity, p=0.004 |
| | Low similarity | 2.47 | High similarity vs. High dissimilarity, p=0.002 |
| | High dissimilarity | 2.46 | Low dissimilarity vs. High dissimilarity, p=0.044 |
| Openness to experience (F=0.897, p=0.442) | High similarity | 2.65 | None |
| | High dissimilarity | 2.59 | |
| | Low similarity | 2.58 | |
| | Low dissimilarity | 2.50 | |